\documentclass[amssymb,aps,twocolumn,showpacs,nofootinbib]{revtex4}
\usepackage[]{graphicx}
\usepackage[]{amsmath}
\usepackage{hyperref}

\def\ket#1{| #1 \rangle}

\def\II{1\!\mathrm{l}}

\def\cB{\mathcal{B}}

\def\cE{\mathcal{E}}
\def\cG{\mathcal{G}}

\def\cL{\mathcal{L}}

\def\cP{\mathcal{P}}

\def\cR{\mathcal{R}}
\def\cS{\mathcal{S}}

\begin{document}

\title{Stabilizer Formalism for Operator Quantum Error Correction}
\author{David Poulin}
\email{dpoulin@iqc.ca}
\affiliation{School of Physical Sciences, The University of Queensland, QLD 4072, Australia}

\date{\today}

\begin{abstract}
Operator quantum error correction is a recently developed theory that provides a generalized framework for active error correction and passive error avoiding schemes. In this paper, we describe these codes in the stabilizer formalism of standard quantum error correction theory. This is achieved by adding a ``gauge" group to the standard stabilizer definition of a code that defines an equivalence class between encoded states. Gauge transformations leave the encoded information unchanged; their effect is absorbed by virtual gauge qubits that do not carry useful information. We illustrate the construction by identifying a gauge symmetry in Shor's 9-qubit code that allows us to remove 4 of its 8 stabilizer generators, leading to a simpler decoding procedure and a wider class of logical operations without affecting its essential properties. This opens the path to possible improvements of the error threshold of fault-tolerant quantum computing. 
\end{abstract}


\maketitle

The theory of fault-tolerant quantum computation \cite{Sho95a,Ste96c,Got97a,KL97a,KLZ98a,Pre99a} demonstrates the formal possibility of efficiently storing and manipulating quantum data for arbitrary long times even in the presence of noise, provided the noise level is bellow a certain threshold. Fault tolerance builds on quantum error correction (QEC) \cite{Sho95a,Ste96c,KL97a,BDSW96a}, which is a mean to actively protect quantum information against noise. Encoded quantum states are restricted to a code subspace $C$ of the system's Hilbert space $H = C \oplus C^\perp$. Measurements are performed to detect if the noise has taken the system out of $C$, and if required, a transformation is applied to restore it. A good code must protect the information against a wide range of errors, and admit simple encoding, error correction procedures, and fault-tolerant gates.

{\em Operator quantum error correction} (OQEC), recently introduced in \cite{KLP05a,KLPL05a}, generalizes the standard theory of QEC and provides a unified framework for active error correction and passive error avoiding techniques such as decoherence-free subspaces \cite{DG97c,ZR97c,LCW98a} and noiseless subsystems \cite{KLV00a,Zan01b,KBLW01a}. In this new paradigm, information is encoded in a subsystem $A$ of the code space $C = A\otimes B$, and errors need only to be corrected modulo a transformation on $B$. The standard QEC theory corresponds to the special case where $B$ is one-dimensional. While this generalization does not lead to new families of codes, it does allow for new error correction procedures, possibly enriching the fault tolerance theory. A prime example is Bacon's OQEC code \cite{Bac05a} that appears to have self-correcting properties. 

Most of the QEC codes  used for fault tolerance constructions can be described with the stabilizer formalism (see Ref.~\cite{Got97a} and references therein). In particular, the first QEC codes proposed by Shor \cite{Sho95a} and Steane \cite{Ste96a} are stabilizer codes. Other important examples include CSS codes \cite{CS96a,Ste96c}, topological codes \cite{Kit03a}, and convolutional codes \cite{OT03a}. A stabilizer formalism has also been constructed to describe the passive error avoiding techniques of decoherence-free subspaces and noiseless subsystems~\cite{KBLW01a}. Aditionally, the stabilizer formalism plays a central role in other branches of quantum information science, e.g. in the so called ``one-time" or ``cluster state" quantum computation model \cite{RB01a}.  Some of the advantages of the stabilizer formalism are that it provides a compact description of QEC codes, admits compact description of a restricted class of dynamical systems (the so-called Clifford group~\cite{Got97a}), and allows to build on classical coding theory (particularly via the CSS construction). 

In this article, we describe a stabilizer formalism for OQEC. We will briefly review the basic theory of OQEC and the standard stabilizer formalism. Then, we demonstrate a general procedure based on the algebraic approach of Ref.~\cite{ZLL04a} to describe the subsystem structure $A\otimes B$ using the Pauli group. We also discuss bounds that apply to these codes. Finally, we illustrate the stabilizer formalism by constructing an OQEC code based on  Shor's 9-qubit code, but which contains a nontrivial $B$ subsystem. This code has all the essential features of Shor's original code, but admits a simpler error recovery procedure and a wider class of encoded operations. 

\smallskip\noindent{\em OQEC theory} --- Let us first summarize the OQEC theory. A fixed partition of the system's Hilbert space $H = A\otimes B \oplus C^\perp$ is assumed. Information is encoded on the $A$ subsystem, i.e. the logical quantum state $\rho^A \in \cB(A)$ is encoded as $\rho^A\otimes\rho^B \oplus 0^{C^\perp}$ with an arbitrary $\rho^B$. We say that the physical map $\cE: \cB(H) \rightarrow \cB(H)$ is {\em correctable} on subsystem $A$ when there exists a physical map $\cR: \cB(H) \rightarrow \cB(H)$ that reverses its action, up to a transformation on the $B$ subsystem, i.e. for all $\rho^A$ and $\rho^B$, $\cR\circ\cE(\rho^A\otimes\rho^B) = \rho^A\otimes\rho^{\prime B}$ for some arbitrary $\rho^{\prime B}$. In terms of the operator-sum representation $\cE(\rho) = \sum_a E_a\rho E_a^\dagger$, the existence of a recovery map $\cR$ requires the following condition to hold (see \cite{KLP05a,KLPL05a})
\begin{equation}
P E_a^\dagger E_b P = \II^A\otimes g_{ab}^B\ \ \forall a,b
\label{eq:gecc}
\end{equation}
where $P$ is the projector onto the code space --- i.e. $PH = C = A\otimes B$ --- and $g_{ab}^B$ is an arbitrary operator in $\cB(B)$. That this condition is also sufficient for $\cE$ to be correctable was proven in \cite{NP05a}, along with alternative information-theoretic necessary and sufficient conditions.  As expected, when the $B$ subsystem is one-dimensional, Eq.~(\ref{eq:gecc}) reduces to the familiar error correction condition~\cite{BDSW96a,KL97a}. 

\smallskip\noindent{\em Stabilizer formalism} --- Let us now focus on the case where the system is composed of $n$ qubits, so $H = \mathbb{C}^{2^n}$. The Pauli matrices are defined as
\begin{equation*}
X = 
\left( \begin{array}{cc}
0 & 1 \\
1 & 0
\end{array} \right),\ \ 
Y = 
\left( \begin{array}{cc}
0 & -i \\
i & 0
\end{array} \right),\ \ 
Z = 
\left( \begin{array}{cc}
1 & 0 \\
0 & -1
\end{array} \right).
\end{equation*}
We denote $X_j$ the matrix $X$ acting on the $j$th qubit, and similarly for $Y_j$ and $Z_j$. The Pauli group on $n$ qubits $\cP_n$ is generated under multiplication by the Pauli matrices acting on each qubits, together with the imaginary number $i$. In terms of independent generators, we have $\cP_n = \langle i,X_1,Z_1,\ldots,X_n,Z_n\rangle$. 

The first step in constructing a stabilizer code is to choose a set of $2n$ operators $\{X_j',Z_j'\}_{j=1,\ldots n}$ from $\cP_n$ that is isomorphic to the set of single qubit Pauli operators $\{X_j,Z_j\}_{j=1,\ldots n}$ in the sense that the primed and unprimed operators obey the same commutation relations among themselves. The operators $\{X_1',Z_j'\}_{j=1,\ldots n}$ generate $\cP_n$ and behave as single-qubit Pauli operators --- we can think of them as acting on $n$ virtual qubits. However, these virtual qubits have no relation whatsoever with the original ``bare" qubits (those related to the {\em unprimed} Pauli operators): the operators $X_j'$ and $Z_j'$ can act non-trivially on several bare qubits, i.e. they are collective degrees of freedom. It is crucial to note that there are several ways of choosing these operators, and that these various choices will lead to different codes. It should therefore be kept in mind that beside the imposed commutation relations, the $X_j'$ and $Z_j'$ are arbitrary.

The stabilizer group $\cS = \langle S_1,\ldots S_{s}\rangle$ with $s\leq n$ is an Abelian subgroup of $\cP_n$ that does not contain $-1$. Without loss of generality, we can choose $S_j = Z_j'$ for $j=1,\ldots, s$. These generators are independent commuting elements of $\cP_n$, so they can be simultaneously diagonalized. The code space $C$ is the span of the vectors fixed by $\cS$, i.e. $S_j \ket\psi = \ket\psi$ for all $j = 1,\ldots s$, and it has dimension $2^{n-s}$. The projector onto the code space is denoted $P$ and obviously satisfies $S_jP = P$ for all $j$. 

The normalizer of $\cS$, denoted $N(\cS)$, is the subgroup of $\cP_n$ that commutes with every element of $\cS$. (This simple definition of the normalizer follows from the fact that every pair of element of $\cP_n$ either commute or anti-commute.) Clearly then, elements of $N(\cS)$ map the code subspace to itself. Given the above construction, we see that $N(\cS) = \langle i,Z'_1,\ldots Z'_n,X'_{s+1},\ldots X'_n \rangle$. 

\smallskip\noindent{\em Stabilizer formalism for OQEC} --- The stabilizer $\cS$ specifies the code subspace $C$, and we must now define a partition of $C$ into subsystems $A\otimes B$. For this, we follow the procedure of Ref.~\cite{ZLL04a} and identify subsystems via the algebra of operators acting on them.\footnote{While we mostly focus on group structures here, an equivalent algebraic description can straightforwardly be obtained by considering the associated group algebras.} Central to the OQEC theory is a notion of equivalence between states: the two states $\rho^A\otimes \rho^B$ and $\rho^A\otimes \rho^{\prime B}$ are considered to carry the same information even if $\rho^B$ and $\rho^{\prime B}$ differ. To capture this notion, we quotient the code state space $\cB(C)$ by a set of ``gauge" transformation $\cG$ that defines an  equivalence relation $\rho \sim \rho' \Leftrightarrow \ \exists g \in \cG:\ \rho = g\rho' g^\dagger$. For $\sim$ to define an equivalence relation, $\cG$ must have a group structure. Clearly, $\cS$ and $i$ should be in $\cG$ as they leave states of $C$ invariant under conjugation. For $\sim$ to keep states in the code subspace, $\cG$ must be a subgroup of $N(\cS)$. Given these properties, $\cG$ is a normal subgroup of $N(\cS)$, so $\cL = N(\cS)/\cG$ also has a group structure (the quotient group). 

The gauge group $\cG \supseteq \{\cS,\langle i\rangle \}$ can thus be generated by the stabilizer generators, the complex number $i$, and an arbitrary subset of the $X_j'$ and $Z'_j$ with $j > s$, i.e. $\cG = \langle i,S_1,\ldots, S_{n-r-k},X_{i_1}',\ldots, X_{i_a}', Z_{j_1}',\ldots, Z_{j_b}'\rangle$ where $\{i_k\}$ and $\{j_k\}$ are subsets of $\{s+1,\ldots n\}$. However, for the two groups $\cG$ and $\cL$ to induce a subsystem structure on $C$, we must have $[\cG,\cL] = 0$ (see \cite{ZLL04a}). As a consequence, the $X_i'$ and $Z_j'$ generators of $\cG$ must always appear in pairs, so without lost of generality, we must have $\cG = \langle i, S_1,\ldots S_s,X_{s+1}',Z_{s+1}',\ldots, X_{s+r}',Z_{s+r}'\rangle$ with $s+r \leq n$. Clearly then, $\cL \simeq \langle X_{s+r+1}',Z_{s+r+1}',\ldots, X_n',Z_n' \rangle$. With a slight abuse of notation, we will henceforth use $\cL$ to denote the quotient group $N(\cS)/\cG$ and its representation on $H$ given above. Since $[\cG,\cL] = 0$ and $\cG \times \cL \simeq N(\cS)$, it follows from Ref.~\cite{ZLL04a} that these groups induce a subsystem structure on the code subspace $C = A\otimes B$, such that the action of any $L\in \cL$ and $g \in \cG$ restricted to the code subspace $C$ is given by
\begin{eqnarray}
gP &=&  \II_{2^k}^A \otimes g^B, {\mathrm{for\ some}} \ \ g^B \in \cB(B), \label{eq:Q}\\
LP &=& L^A \otimes \II^B_{2^{r-k}}, {\mathrm{for\ some}} \ \ L^A \in \cB(A) \label{eq:L}
\end{eqnarray} 
with $A \simeq (\mathbb{C}^2)^{\otimes k}$ and $B\simeq (\mathbb{C}^2)^{\otimes r}$ as desired.

To sum up, we have partitioned the $n$ virtual qubits defined through the $X_j'$ and $Z_j'$ into 3 sets: $s$ {\em stabilizer} qubits, $r$ {\em gauge} qubits, and $k$ {\em logical} qubits, with $s+r+k = n$. The $Z'_j$ operators from the first set are denoted $S_j$ with $j=1,\ldots, s$ respectively. They are stabilizer generators and fix the $2^{r+k}$-dimensional code space $C$. The $Z'_{s+j}$ and $X'_{s+j}$ operators from the second set are denoted $g^z_j$ and $g^x_j$ with $j=1,\ldots, r$. They generate the group $\cL_B$ of  Pauli operations acting on the $r$ virtual qubits of the $B$ subsystem. These qubits do not encode useful information: their sole purpose is to absorb transformations from $\cG$, and as such, they are referred to as gauge qubits. Together with the stabilizer and the complex number $i$, this set generates the gauge group that leaves the encoded information invariant under conjugation, $\cG = \cL_B \times \cS \times \langle i\rangle$. From this definition, it is clear that an Abelian gauge group corresponds to the standard stabilizer formalism, while non-Abelian $\cG$ yield OQEC codes. Finally the $Z'_{s+r+j}$ and $X'_{s+r+j}$ operators from the third set are denoted $\overline Z_j$ and $\overline X_j$ with $j=1,\ldots,k$ respectively. They generate the logical operations $\cL$, and act only on the $k$ virtual qubits of the $A$ subsystem.

Although we have given an explicit set of generators for $\cL$, we stress that only the coset structure of $\cL$ really matters. Operations related by a gauge transformation have the same effect on the encoded qubits, e.g any $\overline Z_j' = g\overline Z_j$ with $g \in \cG$ can serve as the logical $Z$ Pauli operator acting on the $j$th encoded qubit. This defines an equivalence relation $Z \sim Z' \Leftrightarrow ZZ' \in \cG$ between quantum operations. As mentioned in footnote 1, the definitions of the gauge group $\cG$ and logical operations $\cL$ can be extended by considering the associated group algebras: any linear combination of elements of $\cG$ (resp. $\cL$) is an operator acting solely on the gauge system $B$ (resp. encoded qubits $A$). This extends the notion of equivalence relations between states and operations in an obvious way.

\smallskip\noindent{\em Error correction} --- We now study the effect of a set of errors $\{E_a\} \subset \cP_n$. Although this may appear restrictive, we note that a recovery procedure $\cR$ that corrects $\{E_a\}$ will also correct any set of errors obtained from linear combination of elements of $\{E_a\}$. Error detection is made by measuring the the stabilizer generators $S_1,\ldots,S_s$. These give a set of outcomes $(m_1,\ldots,m_s)$ taking values $\pm 1$, called the error syndrome. The all-ones syndrome indicates that the state is in the code subspace $C$, while any other syndrome indicate that an error has taken the state out of $C$. Thus, {\em detectable} errors are those that anti-commute with at least one of the stabilizer generator or that have no effect on the encoded data.

To be {\em correctable}, the set of errors must satisfy Eq.~(\ref{eq:gecc}). [Note that the conjugation ($\dagger$) is irrelevant here, so we will omit it.]  For any pair $a,b$, the operator $E_aE_b$ is an element of either $\cP_n - N(\cS)$, or $N(\cS)-\cG$, or $\cG$. In the first case, there exists an $S \in \cS$ for which $\{E_aE_b,S\} = 0$. Plugging in to Eq.~(\ref{eq:gecc}), we get $PE_aE_bP = PE_aE_bSP = -PSE_aE_bP = -PE_aE_bP = 0$, so the operator error correction condition is fulfilled. In the second case, observe that $N(\cS) - \cG \simeq \{\cL - \II\} \times \cG$, so Eqs.~(\ref{eq:L},\ref{eq:Q}) show that $PE_aE_bP = L^A_{ab}\otimes g^B_{ab}$ for some $L^A_{ab} \neq \II^A$, so these errors cannot be corrected. For the third case, Eq.~(\ref{eq:Q}) shows that $PE_aE_bP = \II^A \otimes g_{ab}^B$, so the condition is satisfied. Therefore, $\{E_a\}$ is a correctable set of errors if and only if $E_aE_b \notin N(\cS) - \cG$ for all pairs $a,b$. 

To construct the recovery procedure, observe that equivalent errors $E_a \sim E_b$ have by definition and Eq.~(\ref{eq:Q}) $PE_aE_bP = \II^A \otimes g_{ab}^B$, and yield the same error syndrome: for all $S \in \cS$ and $g \in \cG$, $[gE_a,S] = 0$ if and only if $[E_a,S]=0$. Thus, syndrome measurement can identify the coset of $\{E_a\}/\cG$ to which the error that occurred belongs. To recover the information encoded in $A$, we can apply any element of that coset to the state. The overall effect of this procedure will be a gauge transformation since equivalent errors have $E_aE_bP = \II^A\otimes g_{ab}^B$ by virtue of Eq.~(\ref{eq:Q}), leaving the logical qubits $A$ unaffected. 

\smallskip\noindent{\em Bounds} --- The distance $d$ of a code is given by the minimal weight\footnote{The weight of an element of $\cP_n$ is the number of its single-qubit Pauli operators in it that differ from the identity.} of operators in $N(\cS) - \cG$. A code of distance $d$ can correct errors on up to $(d-1)/2$ qubits. A stabilizer OQEC code therefore has $4$ parameters, $[[n,k,r,d]]$ representing respectively the number of physical qubits, the number of encoded logical qubits, the number of gauge qubits, and the distance of the code. The Knill-Laflamme or quantum Singleton bound $n \geq 2(d-1) + k$ restricts the possible values of these parameters~\cite{KL97a}. As any bound relating $n$, $k$, and $d$ derived in the context of stabilizer QEC, this bound also applies to OQEC. This follows straightforwardly from Theorem 3 of Ref.~\cite{KLP05a}. Indeed, a $[[n,k,r,d]]$ OQEC code can be transformed into a $[[n,k,0,d]]$ QEC code by turning the gauge $Z$ operators $g^z_j$ into extra stabilizer generators, i.e. by fixing the gauge. 

The theory of OQEC opens the possibility of simplifying existing codes --- turning a $[[n,k,0,d]]$-code into a $[[n,k,r,d]]$-code with $r>0$ --- by identifying ``gauge symmetries" in their stabilizer. This would lead to more efficient error correction procedures with less error syndromes to measure and wider class of encoded operations. A specific example is presented in the next section. The key task is thus to find the largest value of $r$ achievable given values of $n$, $k$, and $d$. We have not yet derived a general bound for the number of gauge qubits beside the trivial observation that at least one stabilizer must be measured when $d > 0$. By exhaustive search however, we have ruled out the existence of a ``better than perfect" quantum code --- a 5-qubit code protecting one logical qubit against  any single qubit error~\cite{LMPZ96a,BDSW96a}, but which requires less than 4 stabilizer generators, i.e. that admits one gauge qubit.

\smallskip\noindent{\em Example} ---  Let us illustrate the idea of reducing the number of stabilizer generators by identifying gauge symmetries using Shor's $[[9,1,0,3]]$ code~\cite{Sho95a}. The stabilizer generators and encoded Pauli operators for this code are given in Table~\ref{t:Shor}.

\begin{table}[]
\begin{tabular}{l|ccccccccc}
\hline\hline
$S_1$ & $X$ & $X$ & $X$ & $X$ & $X$ & $X$ & $\II$ & $\II$ & $\II$\\
$S_2$ & $X$ & $X$ & $X$ & $\II$ & $\II$ & $\II$ & $X$ & $X$ & $X$\\
$S_3$ & $Z$ & $Z$ & $\II$ & $\II$ & $\II$ & $\II$ & $\II$ & $\II$ & $\II$\\
$S_4$ & $\II$ & $Z$ & $Z$ & $\II$ & $\II$ & $\II$ & $\II$ & $\II$ & $\II$\\
$S_5$ & $\II$ & $\II$ & $\II$ & $Z$ & $Z$ & $\II$ & $\II$ & $\II$ & $\II$\\
$S_6$ & $\II$ & $\II$ & $\II$ & $\II$ & $Z$ & $Z$ & $\II$ & $\II$ & $\II$\\
$S_7$ & $\II$ & $\II$ & $\II$ & $\II$ & $\II$ & $\II$ & $Z$ & $Z$ & $\II$\\
$S_8$ & $\II$ & $\II$ & $\II$ & $\II$ & $\II$ & $\II$ & $\II$ & $Z$ & $Z$\\
\hline
$\overline Z $ & $ Z$ & $Z$ & $Z$ & $Z$ & $Z$ & $Z$ & $Z$ & $Z$ & $Z$\\
$\overline X $ & $ X$ & $X$ & $X$ & $X$ & $X$ & $X$ & $X$ & $X$ & $X$\\
\hline\hline
\end{tabular}
\caption{Stabilizer generators and encoded Pauli's for Shor's $[[9,1,0,3]]$ code.}
\label{t:Shor}
\end{table}

\begin{table}[]
\begin{tabular}{l|ccccccccc}
\hline\hline
$S_1$ & $X$ & $X$ & $X$ & $X$ & $X$ & $X$ & $\II$ & $\II$ & $\II$\\
$S_2$ & $X$ & $X$ & $X$ & $\II$ & $\II$ & $\II$ & $X$ & $X$ & $X$\\
$S_3'$ & $Z$ & $Z$ & $\II$ & $Z$ & $Z$ & $\II$ & $Z$ & $Z$ & $\II$\\
$S_4'$ & $\II$ & $Z$ & $Z$ & $\II$ & $Z$ & $Z$ & $\II$ & $Z$ & $Z$\\
\hline
$\overline Z $ & $ Z$ & $Z$ & $Z$ & $Z$ & $Z$ & $Z$ & $Z$ & $Z$ & $Z$\\
$\overline X $ & $ X$ & $X$ & $X$ & $X$ & $X$ & $X$ & $X$ & $X$ & $X$\\
\hline
$g^z_1$ & $\II$ & $Z$ & $Z$ & $\II$ & $\II$ & $\II$ & $\II$ & $\II$ & $\II$\\
$g^x_1$ & $\II$ & $\II$ & $X$ & $\II$ & $\II$ & $\II$ & $\II$ & $\II$ & $X$\\
$g^z_2$ & $\II$ & $\II$ & $\II$ & $\II$ & $Z$ & $Z$ & $\II$ & $\II$ & $\II$\\
$g^x_2$ & $\II$ & $\II$ & $\II$ & $\II$ & $\II$ & $X$ & $\II$ & $\II$ & $X$\\
$g^z_3$ & $Z$ & $Z$ & $\II$ & $\II$ & $\II$ & $\II$ & $\II$ & $\II$ & $\II$\\
$g^x_3$ & $X$ & $\II$ & $\II$ & $\II$ & $\II$ & $\II$ & $X$ & $\II$ & $\II$\\
$g^z_4$ & $\II$ & $\II$ & $\II$ & $Z$ & $Z$ & $\II$ & $\II$ & $\II$ & $\II$\\
$g^x_4$ & $\II$ & $\II$ & $\II$ & $X$ & $\II$ & $\II$ & $X$ & $\II$ & $\II$\\
\hline\hline
\end{tabular}
\caption{Stabilizer generators, encoded Pauli's, and generators of $\cL_B$ for a $[[9,1,3,3]]$ version of Shor's code.}
\label{t:Shor'}
\end{table}

By inspection, we see that it is possible to combine the last $6$ stabilizers of this code, thus eliminating $4$ of them. The remaining $4$ stabilizers hence define a $2^5$ dimensional code space, i.e. $C$ contains $5$ virtual qubits. However, these $5$ qubits are not protected against all single qubit errors; only one logical qubits is immune to noise, while the other $4$ extra qubits in $C$ are gauge qubits. Indeed, the code defined by Table~\ref{t:Shor'} is a $[[9,1,4,3]]$ code, and it coincides with the OQEC code defined in \cite{Bac05a}. This new code has all the essential features of the original code. In particular, it protects one qubit of information against any single qubit error, and it has all the features of a CSS code (e.g. fault-tolerant transversal c-not). It has however lost its ability to protect the logical qubit against some 2-qubit errors, but this is not essential to achieve fault tolerance by concatenation. Note also that there is much more freedom in choosing the encoded operation, e.g. the operator $\II_1 \II_2 X_3\II_4 \II_5\II_6X_7 X_8\II_9 = S_1g^x_1\overline X$ is a valid logical $X$ operation. One can easily verify that the generators $g_k$ of the gauge group generate $\cP_4$, so this code has $4$ gauge qubits as claimed. 

We stress that this code is not the $5$-qubits code \cite{LMPZ96a,BDSW96a} or Steane's 7-qubit code \cite{Ste96a} disguised in a 9-qubit code. Indeed, the stabilizer of this new code is a subgroup of the stabilizer of the original code, and the encoded operations are the same as those of the original code. By exhaustive search, we have established that the 5-qubit code and Steane's 7-qubit code have no gauge symmetry at all. 

\smallskip\noindent{\em Conclusion} --- Operator quantum  error correction theory provides a generalized and unified framework for active error correction techniques and passive error avoiding methods. In this paper, we have developed a stabilizer description of such codes. Stabilizer codes have been central to fault-tolerant constructions, as well as other areas of quantum information science: it is our hope the  generalization presented here will enrich these subjects. We have demonstrated that bounds which restrict the families of achievable codes derived in the setting of standard stabilizer QEC theory apply straightforwardly to OQEC codes via gauge fixing. Finally, we have illustrated our formalism by identifying ``gauge" symmetries in Shor's code that lead to substantial simplifications. An important issue which remains open is to bound the number of gauge qubits that can be identified given the other parameters of the code.

\smallskip\noindent{\em Acknowledgments} --- We thank Michael Nielsen for stimulating discussions on the present topic, and Andreas Klappenecker and Dave Bacon for pointing out errors in a previous version of this paper. 


\end{document}